# Longitudinal transport spin polarization of spin degenerate antiferromagnets

Meng Zhu[1,2†], Jianting Dong[1†], Xinlu Li[1,2], Jiahao Shentu[1], Yizhuo Song[1], Evgeny Y. Tsymbal[3*] and Jia Zhang[1*]

[1]*School of Physics and Wuhan National High Magnetic Field Center,*
*Huazhong University of Science and Technology, 430074 Wuhan, China*

[2]*College of Physics and Electronic Science, Hubei Normal University, 435002 Huangshi, China*

[3]*Department of Physics and Astronomy & Nebraska Center for Materials and Nanoscience,*
*University of Nebraska, Lincoln, Nebraska 68588, USA*

[*]tsymbal@unl.edu
[*] jiazhang@hust.edu.cn
[†]Authors contributed equally to this work.

## Abstract

A vital goal in spintronics is the efficient electrical generation of spin currents, a pursuit that has recently focused on using antiferromagnets (AFMs) as spin current sources. It has been demonstrated that antiferromagnets with broken *PT* symmetry (parity + time reversal) can efficiently generate longitudinal and transverse spin currents. At the same time, it has been generally thought that antiferromagnets with *PT* symmetry (*PT*-AFMs) forbid the longitudinal spin polarization due to their spin-degenerate band structure. Here, in contrast to this common expectation, we show, using theoretical analysis based on magnetic point group symmetry, that most *PT*-AFMs can generate longitudinal spin currents due to spin-orbit coupling. Using density-functional theory, we calculate the longitudinal spin conductivity of representative *PT*-AFMs, L1$_0$-MnPt and Mn$_2$Au, and show that its magnitude is comparable to that of their *PT*-broken counterparts. Our symmetry-enabled classification of antiferromagnets and theoretical results for the longitudinal spin conductivity in representative *PT*-AFMs expands our understanding of spin transport and shows the possibility of robust spin-current generation in a broad range of spin-degenerate antiferromagnets.

*Introduction.* Spin current is a central topic in the field of spintronics, offering pathways for the manipulation of magnetic orders in condensed matter [1–4]. The generation, control, and detection of spin currents have led to significant advancements in designing novel spintronic devices, with promising perspectives for high-density memory and nonvolatile logic applications [5,6]. Over the past decades, extensive research has focused on various physical mechanisms that can generate spin currents. The electric field generated transverse spin current, driven by the spin Hall effect, has been widely studied in nonmagnetic, ferromagnetic, and antiferromagnetic materials [7,8]. Especially, the generation of spin current in antiferromagnets (AFMs) has attracted increasing attention, owing to their advantages such as insensitivity to magnetic perturbations, absence of stray fields, and ultrafast spin dynamics [9–11].

The longitudinal spin currents have been well recognized in ferromagnetic metals with their exchange-split electronic structure [12] and spin-split AFMs with their spin polarization in momentum space [13–17]. Especially, spin-split AFMs with broken $PT\tau$ and $U\tau$ symmetries (where $P$ is spatial inversion, $T$ is time reversal, and $\tau$ is fractional lattice translation) could generate longitudinal spin-transport phenomena such as the tunneling magnetoresistance (TMR) effect [18,19]. However, the spin-split AFMs are largely restricted by the limited material candidates for applications. Conventional AFMs with $PT\tau$ symmetry ($PT$-AFMs), and hence spin degenerate band structures, are abundant but whether the longitudinal spin current can exist in these magnets and how it correlates with the antiferromagnetic order parameter remain unclear. In contrast to nonmagnets that rely solely on crystal symmetry [14,15], AFMs offer the spin degree of freedom to manipulate spin currents by the orientation of the Néel vector.

Here, we provide a symmetry-based analysis of the longitudinal spin conductivity in AFMs for all 122 magnetic point groups (MPGs), with emphasis on $PT$-AFMs. We find, based on the features of longitudinal spin conductivity enforced by symmetry, that the MPGs of AFMs can be categorized into five classes. Among them are two classes of $PT$-AFMs supporting longitudinal spin conductivity due to spin-orbit coupling (SOC), which is the main result of this work. By employing first-principles calculations, we demonstrate sizable longitudinal spin conductivity in two representative *PT*-AFMs, L1$_0$-MnPt and Mn$_2$Au. The possibility to generate longitudinal spin currents in *PT*-AFMs greatly broadens the variety of AFMs that can be exploited as a robust spin current source for spintronic applications.

*Classification of longitudinal spin currents in AFMs.* Depending on the presence of *Tτ* and *PTτ* symmetries, as well as *T*-symmetry of longitudinal spin conductivity, different antiferromagnets described by 122 MPGs can be classified into five categories, as illustrated in Fig. 1 and summarized in Table I.

Type-I AFMs support *Tτ* symmetry but break *PTτ* symmetry. In non-centrosymmetric systems, SOC splits the band structure and enables *T*-even longitudinal spin conductivity $\sigma_{ii}^{k}$, where the spin current is parallel to the electric field direction *i*, and the spin polarization is oriented along the *k* direction. It is worth noting that a subset of non-collinear odd-parity ("*p*-wave") AFMs in the same class can even exhibit spin-split bands without SOC [20]. A representative Type-I AFM is MnS$_2$ [21] which belongs to MPG $mm21'$.

Type-II AFMs break both *Tτ* and *PTτ* symmetries and exhibit nonrelativistic spin-splitting. They have spin-split band structures and nonrelativistic *T*-odd longitudinal spin conductivity along certain directions [18]. Representative Type-II AFMs include collinear AFMs, such as altermagnets RuO$_2$ (MPG $4'/mm'm$) and CrSb (MPG $6'/m'mm'$) [22,23], and noncollinear AFMs, such as Mn$_3$Pt (MPG $\bar{3}m'$), Mn$_3$Sn (MPG $m'm'm$), and Mn$_3$GaN (MPG $3m'$) [5,24,25].

Type-III AFMs support both *Tτ* and *PTτ* symmetries. Typical materials in this category include L1$_0$-MnPt [26], Cr [27], and *α*-FeRh [28]. Type-IV AFMs are non-centrosymmetric, violating *Tτ* symmetry but supporting *PTτ* symmetry, and are exemplified by CuMnAs [29] and Mn$_2$Au [30] ( MPG $m'mm$ ). All *PT*-symmetric AFMs exhibit spin-degenerate band structures.

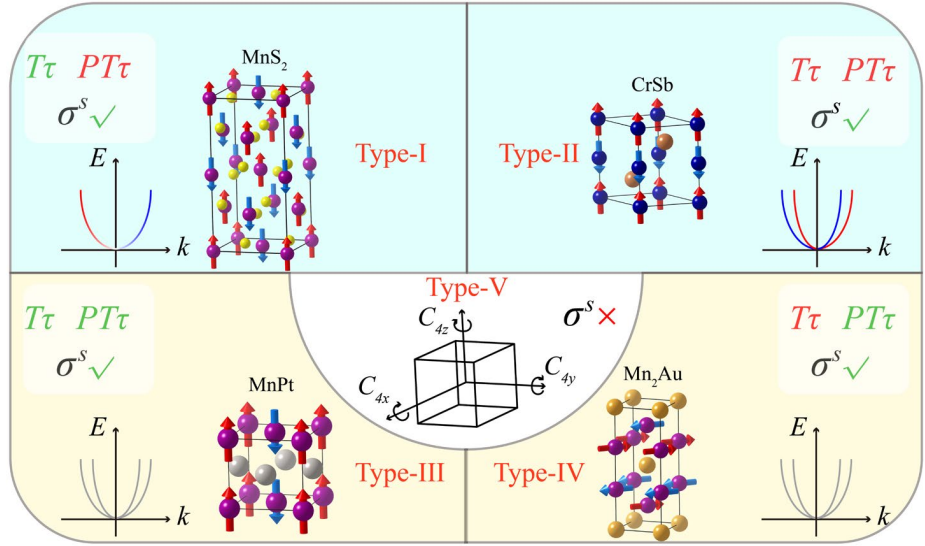

**Figure 1.** Schematic illustration of five classes of antiferromagnets according to $T\tau$ and $PT\tau$ symmetries and longitudinal spin conductivity $\sigma^s$.

AFMs with $PT\tau$ symmetry can also be classified according to their longitudinal spin conductivity, which can be zero or non-zero depending on other symmetries. It is known that spin conductivity tensor contains time reversal odd (*T*-odd) and time reversal even (*T*-even) components [31]. The *T*-odd component changes sign when magnetic moments are reversed and hence, in the presence of $PT\tau$ symmetry, must vanish. On the contrary, the *T*-even component does not change sign when magnetic moments are reversed, and hence the $PT\tau$ symmetry does not necessarily imply that spin conductivity vanishes. There is, however, a subset of *PT*-AFMs where the *T*-even part of longitudinal spin conductivity is zero by symmetry. We categorize this subset of Type-III and IV AFMs as Type-V depicted in Fig. 1. Specifically, Type-V AFMs contain either combination of three two-fold rotational axes ($C_{2[110]}$, $C_{2[011]}$, $C_{2[101]}$) or three four-fold rotation axes ($C_{4[100]}$, $C_{4[010]}$, $C_{4[001]}$), representing MPGs $m\bar{3}m1'$, $m'\bar{3}'m$, and $m'\bar{3}'m'$. These symmetries make the *T*-even part of the spin conductivity tensor to exhibit a specific form where $\sigma_{ii}^{k}$ components vanish:

$$\sigma^x = \begin{pmatrix} 0 & 0 & 0 \\ 0 & 0 & \sigma^x_{yz} \\ 0 & -\sigma^x_{yz} & 0 \end{pmatrix}, \quad \sigma^y = \begin{pmatrix} 0 & 0 & -\sigma^x_{yz} \\ 0 & 0 & 0 \\ \sigma^x_{yz} & 0 & 0 \end{pmatrix}, \quad \sigma^z = \begin{pmatrix} 0 & \sigma^x_{yz} & 0 \\ -\sigma^x_{yz} & 0 & 0 \\ 0 & 0 & 0 \end{pmatrix}.$$

AFMs, which belong to these three MPGs, cannot generate longitudinal spin currents in whatever low-symmetric crystal planes [32]. A representative example of such AFM is NdZn [33] which belongs to MPG $m\bar{3}m1'$.

TABLE I. Classification of 122 MPGs describing AFMs into five classes based on $T$ and $PT$ magnetic point group symmetries (corresponding to $T\tau$ and $PT\tau$ magnetic space group symmetries) and longitudinal spin conductivity classified as $T$-odd or $T$-even. The first 31 MPGs within the Type-II class compatible with ferromagnetic order are listed at the beginning of the class.

| Type | Magnetic point group | $T$ | $PT$ | Band structure | Longitudinal spin conductivity | Examples |
|---|---|---|---|---|---|---|
| I | $11', 21', m1', 2221', mm21', 41', \bar{4}1',$ $4221', 4mm1', \bar{4}2m1', 31', 321', 3m1',$ $61', \bar{6}1', 6221', 6mm1', \bar{6}m21', 231',$ $4321', \bar{4}3m1'$ | ✓ | × | Spin-split | SOC induced, $T$-even | MnS$_2$ |
| II | $1, \bar{1}, 2, 2', m, m', 2/m, 2'/m', 2'2'2,$ $m'm2', m'm'2, m'm'm, 4, \bar{4}, 4/m,$ $42'2', 4m'm', \bar{4}2'm', 4/mm'm', 3, \bar{3},$ $32', 3m', \bar{3}m', 6, \bar{6}, 6/m, 62'2', 6m'm',$ $\bar{6}m'2', 6/mm'm', 222, mm2, mmm, 4',$ $\bar{4}', 4'/m, 422, 4'22', 4mm, 4'm'm,$ $\bar{4}2m, \bar{4}'2'm, \bar{4}'2m', 4/mmm, 4'/mm'm,$ $32, 3m, \bar{3}m, 6', \bar{6}', 6'/m', 622, 6'22',$ $6mm, 6'mm', \bar{6}m2, \bar{6}'m'2, \bar{6}'m2',$ $6/mmm, 6'/m'mm', 23, m\bar{3}, 432, 4'32',$ $\bar{4}3m, \bar{4}'3m', m\bar{3}m, m\bar{3}m'$ | × | × | Non-relativistic spin-split | $T$-odd without SOC; $T$-odd and $T$-even with SOC | CrSb |
| III | $\bar{1}1', 2/m1', mmm1', 4/m1', 4/mmm1',$ $\bar{3}1', \bar{3}m1', 6/m1', 6/mmm1', m\bar{3}1'$ | ✓ | ✓ | Spin-degenerate | SOC induced, $T$-even | L1$_0$-MnPt |
| IV | $\bar{1}', 2'/m, 2/m', m'mm, m'm'm', 4/m',$ $4'/m', 4/m'mm, 4'/m'm'm, 4/m'm'm',$ $\bar{3}', \bar{3}'m, \bar{3}'m', 6'/m, 6/m', 6/m'mm,$ $6'/mmm', 6/m'm'm', m'\bar{3}'$ | × | ✓ | Spin-degenerate | SOC induced, $T$-even | Mn$_2$Au |
| V | $m\bar{3}m1'$, $m'\bar{3}'m$, $m'\bar{3}'m'$ | | -- | ✓ | Spin-degenerate | vanishing | NdZn |

*Symmetry arguments.* Now, we exclude Type-V and focus on Type-III or Type-IV *PT*-AFMs which belong to 29 MPGs and exhibit longitudinal spin currents. These 29 MPGs can be further categorized into two subgroups based on whether their longitudinal spin conductivity vanishes or not when the electric field is applied along a certain direction. We find that *PT*-AFMs of 13 MPGs, namely $mmm1'$, $m'mm$, $m'm'm'$, $4/mmm1'$, $4/m'mm$, $4'/m'm'm$, $4/m'm'm'$, $6/mmm1'$, $6/m'mm$, $6'/mmm'$, $6/m'm'm'$, $m'\bar{3}'$ and $m\bar{3}1'$, cannot produce longitudinal spin currents when the electric field is applied along specific directions. We dub this subgroup "directional *PT*-AFMs." In contrast, *PT*-AFMs of the other 16 MPGs, i.e., $\bar{1}1'$, $\bar{1}'$, $2/m1'$, $2'/m$, $2/m'$, $4/m1'$, $4/m'$, $\bar{3}1'$, $\bar{3}'$, $\bar{3}m1'$, $\bar{3}'m$, $\bar{3}'m'$, $6/m1'$, $6'/m$ and $6/m'$, can always generate longitudinal spin current regardless of the electric field direction. We dub this subgroup "omnidirectional *PT*-AFMs."

The underlying mechanism for directional *PT*-AFMs lies in the symmetry constraints of their MPGs: $C_{2x}$ rotational symmetry eliminates the longitudinal spin conductivity $\sigma_{ii}^y$ and $\sigma_{ii}^z$. Similarly, $C_{2y}$ and $C_{2z}$ rotational symmetries eliminate the longitudinal spin conductivities $\sigma_{ii}^k$, where $k = x, z$ and $k = x, y$, respectively. As a result, when incorporating the electric field into the symmetry analysis, a system with at least two perpendicular twofold rotational symmetries (e.g., a combination of $C_{2x}$ and $C_{2y}$) cannot generate a longitudinal spin current. A longitudinal spin current can only emerge when the electric field is applied along a direction which leaves no more than one rotational symmetry unbroken.

*Case study of a type-III AFM: $L1_0$-MnPt.* Next, we consider representative Type-III AFMs, namely $L1_0$ Mn$X$ ($X$ = Pt, Ir, Rh, Pd, etc.), which belong to the space group P4/mmm (No. 123), and focus on MnPt, as an example. Experiments show that $L1_0$-MnPt is a metallic collinear antiferromagnet with the Néel temperature of around 970 K [34]. The magnetic moments of bulk

L1$_0$-MnPt are aligned along the [001] crystallographic directions (referred to as MnPt$_{[001]}$), as illustrated in Fig. 2(a). The easy axis of L1$_0$-MnPt can also be set to [110] crystallographic axes (referred to as MnPt$_{[110]}$) when it is grown on MgO substrate [35], as depicted in Fig. 2(b). The MPGs of MnPt$_{[001]}$ and MnPt$_{[110]}$ are $4/mmm1'$ (No. 15.2.54) and $mmm1'$ (No. 8.2.25), respectively. Both MPGs contain $PT$ and $T$ symmetries. In addition, the MPG of MnPt$_{[001]}$ is the same as nonmagnetic L1$_0$-MnPt.

Fig. 2 shows the calculated band structures of MnPt$_{[001]}$ and MnPt$_{[110]}$. Due to the presence of $PT$ symmetry, the band structures of MnPt$_{[001]}$ and MnPt$_{[110]}$ are spin degenerate. Compared to MnPt$_{[001]}$, the band structure of MnPt$_{[110]}$ shows anisotropy around the $\Gamma$ point. Importantly, the spin-degenerate band structures do not necessarily imply that the longitudinal spin current is zero, as we discuss next.

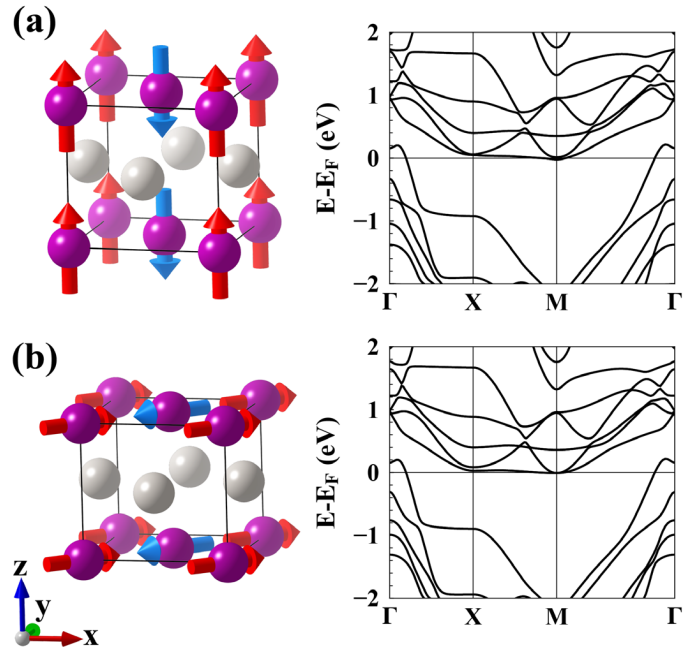

**Figure 2.** Crystal structure and antiferromagnetic spin configuration of (a) L1$_0$-MnPt$_{[001]}$ and (b) MnPt$_{[110]}$ within the (001) crystal facet orientation, as well as their band structures calculated in the presence of SOC. The horizontal black lines indicate the Fermi energy.

Using linear response symmetry analysis [36], we obtain spin conductivity tensors for MnPt[001] and MnPt[110] with the (001) crystal facet orientation (Supplementary Table S1). For both systems, the non-zero spin conductivity elements are produced by SOC and vanish when SOC is not included. MnPt[001], where the Néel vector is pointing along the $z$-direction and thus preserves the four-fold rotational symmetry $C_{4z}$, has the same MPG as its non-magnetic counterparts, like CuAu (MPG $4/mmm1'$) [37]. Its spin conductivity has only non-vanishing off-diagonal terms, i.e., spin Hall conductivity, while the diagonal terms, i.e., the longitudinal spin conductivity, are zero. In contrast, in MnPt[110], the Néel vector is pointing along the [110] direction, which breaks the $C_{4z}$ symmetry and leads to non-zero spin conductivity components. As seen from Table S1, MnPt[110] has finite longitudinal spin conductivities $\sigma_{xx}^z$ and $\sigma_{yy}^z$, which indicate that the $z$-polarized longitudinal spin current can be generated for the electric field applied along $x$ or $y$ axis.

It is evident that the longitudinal spin conductivities of MnPt[110] are even under time reversal. For instance, mirror reflection $M_z$ reverses the magnetic moments of Mn atoms and the Néel vector, while does not alter the applied electric field and the longitudinal spin current with the $z$ polarization. We calculate the longitudinal spin conductivity of MnPt[110] by performing full-relativistic first-principles calculations (see Supplementary Note 1 [38]). We find that the longitudinal spin conductivity is remarkably large with $\sigma_{xx}^z = -\sigma_{yy}^z = 276\,\hbar/2e\,(\Omega\cdot\text{cm})^{-1}$, which is on the same order as the reported values for $PT$-broken AFMs, such as Mn$_3$Pt (100 $\hbar/2e\,\Omega^{-1}\text{cm}^{-1}$) [39] and RuO$_2$ (114 $\hbar/2e\,\Omega^{-1}\text{cm}^{-1}$) [40]. Note that due to the presence of two-fold rotation $C_{2[110]}$, $\sigma_{xx}^z$ and $\sigma_{yy}^z$ must have equal magnitudes but opposite sign.

The emergence of longitudinal spin currents in MnPt[110] can be further understood by examining the constraints on the diagonal spin conductivities imposed by each individual MPG operation. As listed and compared in Supplementary Table S2, both two-fold rotation $C_{2z}$ and mirror symmetry $M_z$ forbid longitudinal spin currents polarized along $x$ and $y$ directions but support spin currents with $z$-polarization. Yet, the $\sigma_{zz}^z$ element is prohibited by two-fold rotation operations and mirror operations along [$\bar{1}10$] ($C_{2[\bar{1}10]}$ and $M_{[\bar{1}10]}$), which further reduce the surviving diagonal elements to contain only $\sigma_{xx}^z$ and $\sigma_{yy}^z$.

*Anisotropy of longitudinal spin conductivity.* In addition to its large magnitude, the longitudinal spin conductivity of *PT*-AFMs also exhibits pronounced anisotropy. For MnPt[110], when the electrical field $E$ is applied in the *xy* plane with angle $\theta$ away from the [100] crystal axis (Fig. 3a), longitudinal spin conductivities $\sigma_{xx}^z(\theta)$ and $\sigma_{yy}^z(\theta)$ can be expressed as (see Supplementary Note 4 for the derivation):

$$\sigma_{xx}^z(\theta) = \sigma_{xx}^z(0)\cos(2\theta) \qquad \sigma_{yy}^z(\theta) = -\sigma_{xx}^z(0)\cos(2\theta). \qquad (1)$$

The calculated longitudinal spin conductivities of MnPt[110] as a function of angle $\theta$ are shown in Fig. 3(a). It is seen that $\sigma_{xx}^z$ is highly anisotropic and follows a cosine function with 180° periodicity. The maximum longitudinal spin conductivity $\sigma_{xx}^z$ is $276\,\hbar/2e\,(\Omega\cdot\text{cm})^{-1}$ and $-276\,\hbar/2e\,(\Omega\cdot\text{cm})^{-1}$ when $E$ is along the *x* and *y* axis, respectively. When the applied field is rotated by 90°, the longitudinal spin conductivity flips sign.

Although the longitudinal spin current is absent in MnPt[001] when the electric field is applied in the (001) plane, MnPt[001] can still produce a longitudinal spin polarization when the electric field is applied within certain low-symmetry crystal planes. This is due to lifting the fourfold rotation symmetry $C_{4z}$ and the anisotropic spin Hall effect. In particular, for the low symmetry (010) plane of MnPt[001], there are diagonal spin conductivity elements $\sigma_{xx}^y$ and $\sigma_{zz}^y$, which arise from the anisotropic spin Hall terms satisfying the condition $\sigma_{xz}^y \neq -\sigma_{zx}^y$. As a result, when the electric field is applied in the (010) plane with angle $\theta$ between the *x* axis, MnPt[001] can also have non-zero spin conductivity $\sigma_{xx}^y(\theta)$ and $\sigma_{zz}^y(\theta)$ (see Supplementary Note 4 for derivation):

$$\sigma_{xx}^y(\theta) = \frac{1}{2}[\sigma_{xz}^y(0)+\sigma_{zx}^y(0)]\sin(2\theta) \qquad \sigma_{zz}^y(\theta) = -\frac{1}{2}[\sigma_{xz}^y(0)+\sigma_{zx}^y(0)]\sin(2\theta). \qquad (2)$$

Fig. 3(c-d) shows anisotropic longitudinal spin conductivity $\sigma_{xx}^{y}(\theta)$ and $\sigma_{zz}^{y}(\theta)$ for MnPt[001] as a function of $\theta$. In the shape of a sinusoidal function, $\sigma_{xx}^{y}(\theta)$ peaks at $\theta = 45°$ with maximum of $86\,\hbar/2e\,(\Omega\cdot\text{cm})^{-1}$. Obviously, both MnPt[110] and MnPt[001] belong to the directional $PT$-AFMs, and their longitudinal spin conductivity vanishes when the electric field is applied along particular high-symmetry directions.

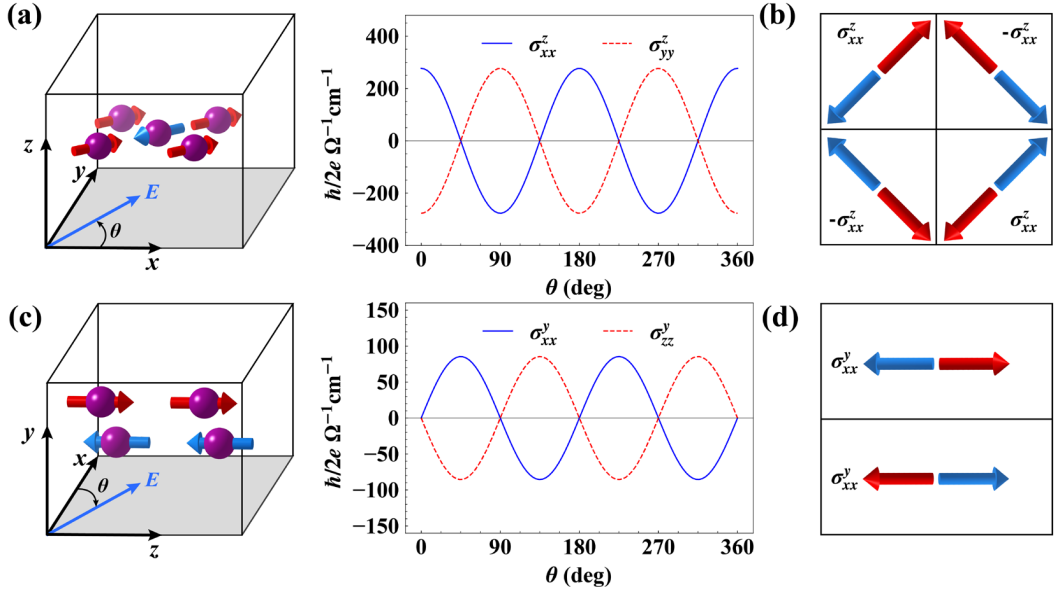

**Figure 3**. (a) Schematic illustration of (001)-oriented MnPt[110] with $\theta$ representing the angle between $x$ axis and the applied electric field in the $xy$-plane. The calculated longitudinal spin conductivity $\sigma_{xx}^{z}$ (blue line) and $\sigma_{yy}^{z}$ (red dashed line) as a function of angle $\theta$. (b) The longitudinal spin conductivity $\sigma_{xx}^{z}$ in four energy degenerate AFM domains, where the red and blue arrows indicate opposite magnetic moments on Mn atoms. (c) The same as (a) but for (010)-oriented MnPt[001] with $\theta$ representing the angle between the $x$ axis and the applied electric field in the $xz$-plane. The calculated anisotropic longitudinal spin conductivity $\sigma_{xx}^{y}$ (blue line) and $\sigma_{zz}^{y}$ (red dashed line) of MnPt[110] as a function of angle $\theta$. (d) Schematic of the longitudinal spin conductivity $\sigma_{xx}^{y}$ in two energy degenerate AFM domains.

Realistic AFMs are often stabilized in multiple energy-degenerate domain states. Especially, for collinear AFMs with high Néel temperature and large exchange interactions, field cooling process in a large magnetic field is required to set the AFM in a single-domain state. We therefore discuss the effect of AFM domains on the longitudinal spin current. In MnPt$_{[110]}$, the Néel vector can be aligned along four directions different by 90°: [110], [1$\bar{1}$0], [$\bar{1}$10] and [$\bar{1}\bar{1}$0]. The AFM domains corresponding to the Néel vector along the [110] and [$\bar{1}\bar{1}$0] directions produce spin currents of opposite sign compared to those associated with the [$\bar{1}$10] and [1$\bar{1}$0] directions. As a result, sign of the longitudinal spin current in MnPt$_{[110]}$ is determined by the AFM domain distribution. In contrast, the Néel vector of AFM domains in MnPt$_{[001]}$ can be pointed along either [001] or [00$\bar{1}$] directions, which are different by 180°. These AFM domains generate the spin current of the same sign since the studied longitudinal spin conductivity is *T*-even. In this case, when the electric field is applied within a low-symmetry crystal plane, such as (010), sign of the longitudinal spin conductivity remains unchanged, and the longitudinal spin current persists across different AFM domains.

*Case study of a type-IV AFM: Mn$_2$Au.* An example of Type-IV AFM is tetragonal Mn$_2$Au [30,41] (MPG $m'mm$) which has high Néel temperature of around 1300 K. As depicted in Fig. 4, two Mn atoms with opposite magnetic moments are connected by space-inversion operation, and therefore Mn$_2$Au breaks *P* symmetry but preserves *PT* symmetry. Thanks to the broken local inversion symmetry of the Mn sites, the Néel spin-orbit torque is present in Mn$_2$Au, which can redirect the Néel vector by applying an electric current [36,42–44]. The calculated band structure of Mn$_2$Au is spin degenerate, as shown in Supplementary Fig. S1. According to the linear response analysis, in the presence of SOC, Mn$_2$Au (001) has two non-zero components of the longitudinal spin conductivity, $\sigma_{xx}^z$ and $\sigma_{yy}^z$, which according to our calculation are equal to $\sigma_{xx}^z = -\sigma_{yy}^z = 113\,\hbar/2e\,(\Omega\cdot\text{cm})^{-1}$ (Supplementary Note 6). Similar to the case of L1$_0$-MnPt, the longitudinal spin conductivity of Mn$_2$Au is anisotropic when the electric field is swept in the *xy* plane. As a result, both the sign and magnitude of the longitudinal spin current in *PT*-AFMs relies on the relative orientation between the Néel vector and the applied electric field, which offers a reliable method for the electrical detection of the Néel vector orientation.

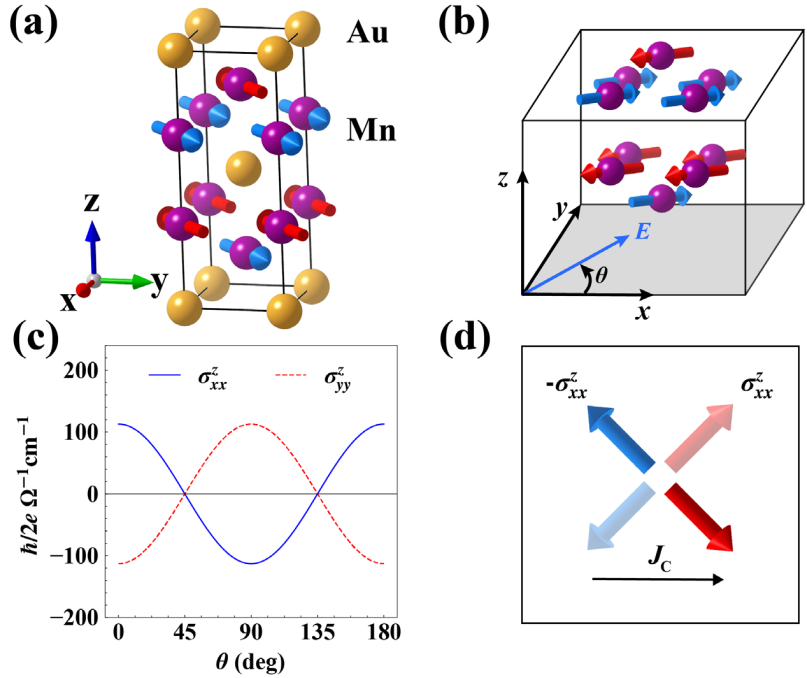

**Figure 4**. (a) Crystal structure and spin configuration of $Mn_2Au$ within the (001) crystal facet orientation. (b) Schematic illustration of $Mn_2Au$ spin configuration with $\theta$ representing the angle between the *x* axis and the applied electric field. (c) Calculated anisotropic longitudinal spin conductivity $\sigma_{xx}^z$ (blue line) and $\sigma_{yy}^z$ (red dashes) of $Mn_2Au$ as functions of rotation angle $\theta$ in the *xy*-plane. (d) Schematic of the longitudinal spin conductivity $\sigma_{xx}^z$ in two energy degenerate AFM domains.

*The implications of longitudinal spin current in PT-AFMs.* Based on the above discussion, the longitudinal spin currents do exist in most *PT*-AFMs, which enable promising spintronics applications. As shown in Fig. 5, the robust and sustainable longitudinal spin current in *PT*-AFMs can be utilized in a tunnel junction to detect the relative magnetic alignment of the electrodes *via* a tunneling magnetoresistance effect and can exert a spin transfer torque on a free layer. In *PT*-AFMs with multiple domains, for instance, $L1_0$-$MnPt_{[110]}$ and $Mn_2Au$, the 90 degrees switching of the Néel vector may also be probed via the TMR effect.

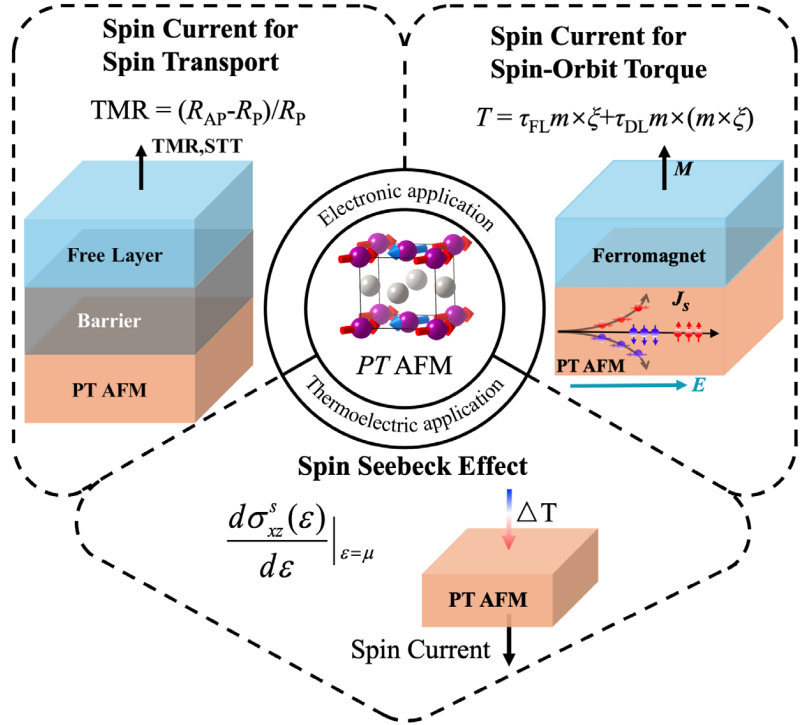

**Figure 5**. Prospective applications of *PT*-AFMs for longitudinal spin transport, spin orbit torque and spin Seebeck effect.

The longitudinal spin currents in *PT*-AFMs can also be useful for other applications. In AFM/FM bilayers, together with a transverse spin current generated by the conventional spin Hall effect, the longitudinal spin currents in *PT*-AFMs can generate nonequilibrium spin density at the interface and exert spin-torque on the adjacent ferromagnet *via* exchange interaction by the Rashba mechanism or inverse spin galvanic effect (ISGE) [2], similar to the SOT in TI(topological insulator)/FM and NM/FM bilayers [45,46].

The discussed electric field driven spin transport phenomena can be naturally extended to thermoelectric applications driven by the temperature gradient. For instance, the longitudinal spin Seebeck effect may be present in *PT*-symmetric AFMs as in ferromagnets [47]. These multifunctional roles highlight the prospective applications of *PT*-AFMs in future spintronic and thermoelectric devices.

In summary, we have predicted the existence of non-vanishing longitudinal spin conductivity in spin degenerate antiferromagnets. We have analyzed the longitudinal spin conductivity of AFMs within all 122 MPGs and categorized them into five classes based on symmetry arguments. By employing the linear response symmetry analysis, the longitudinal spin transport has been verified and demonstrated in collinear antiferromagnets L1$_0$-MnPt and Mn$_2$Au. Using first-principles calculations, we have shown that sizeable *T*-even longitudinal spin conductivity can be generated in *PT*-AFMs when the applied electric field breaks rotational symmetries. Notably, the generated longitudinal spin current is highly anisotropic and robust against multiple AFM domains. Our findings provide theoretical support for the experimental exploration of the longitudinal spin transport in a vast family of spin-degenerate AFMs and useful applications for spintronics.

## Acknowledgement

This work was supported by the National Natural Science Foundation of China (grant No. T2394475, T2394472, T2394470).

## References

[1] S. D. Bader and S. S. P. Parkin, Spintronics, Annu. Rev. Condens. Matter Phys. **1**, 71 (2010).
[2] A. Manchon, J. Železný, I. M. Miron, T. Jungwirth, J. Sinova, A. Thiaville, K. Garello, and P. Gambardella, Current-induced spin-orbit torques in ferromagnetic and antiferromagnetic systems, Rev. Modern Phys. **91**, 035004 (2019).
[3] A. Hirohata, K. Yamada, Y. Nakatani, I.-L. Prejbeanu, B. Diény, P. Pirro, and B. Hillebrands, Review on spintronics: principles and device applications, J. Magn. Magn. Mater. **509**, 166711 (2020).
[4] E. Y. Tsymbal and I. Žutić, Spintronics Handbook: Spin Transport and Magnetism, 2nd ed. (CRC Press, 2019).
[5] P. Qin, H. Yan, X. Wang, H. Chen, Z. Meng, J. Dong, M. Zhu, J. Cai, Z. Feng, X. Zhou, et al., Room-temperature magnetoresistance in an all-antiferromagnetic tunnel junction, Nature **613**, 485 (2023).
[6] C. H. Marrows, J. Barker, T. A. Moore, and T. Moorsom, Neuromorphic computing with

spintronics, Npj Spintron. **2**, 1 (2024).

[7] I. Garate and M. Franz, Inverse spin-galvanic effect in the interface between a topological insulator and a ferromagnet, Phys. Rev. Lett. **104**, 146802 (2010).

[8] J. Sinova, S. O. Valenzuela, J. Wunderlich, C. H. Back, and T. Jungwirth, Spin Hall effects, Rev. Mod. Phys. **87**, 1213 (2015).

[9] H. Chen, L. Liu, X. Zhou, Z. Meng, X. Wang, Z. Duan, G. Zhao, H. Yan, P. Qin, and Z. Liu, Emerging antiferromagnets for spintronics, Adv. Mater. **36**, 2310379 (2024).

[10] P. K. Amiri, C. Phatak, and G. Finocchio, Prospects for antiferromagnetic spintronic devices, Annu. Rev. Mater. Res. **54**, 117 (2024).

[11] V. Baltz, A. Manchon, M. Tsoi, T. Moriyama, T. Ono, and Y. Tserkovnyak, Antiferromagnetic spintronics, Rev. Mod. Phys. **90**, 15005 (2018).

[12] A. Davidson, V. P. Amin, W. S. Aljuaid, P. M. Haney, and X. Fan, Perspectives of electrically generated spin currents in ferromagnetic materials, Phys. Lett. A **384**, 126228 (2020).

[13] J. Krempaský, L. Šmejkal, S. W. D'Souza, M. Hajlaoui, G. Springholz, K. Uhlířová, F. Alarab, P. C. Constantinou, V. Strocov, D. Usanov, et al., Altermagnetic lifting of kramers spin degeneracy, Nature **626**, 517 (2024).

[14] L.-D. Yuan, Z. Wang, J.-W. Luo, and A. Zunger, Prediction of low-Z collinear and noncollinear antiferromagnetic compounds having momentum-dependent spin splitting even without spin-orbit coupling, Phys. Rev. Mater. **5**, 14409 (2021).

[15] J. Dong, X. Li, G. Gurung, M. Zhu, P. Zhang, F. Zheng, E. Y. Tsymbal, and J. Zhang, Tunneling magnetoresistance in noncollinear antiferromagnetic tunnel junctions, Phys. Rev. Lett. **128**, 197201 (2022).

[16] J. Železný, Y. Zhang, C. Felser, and B. Yan, Spin-polarized current in noncollinear antiferromagnets, Phys. Rev. Lett. **119**, 187204 (2017).

[17] G. Gurung, M. Elekhtiar, Q.-Q. Luo, D.-F. Shao, and E. Y. Tsymbal, Nearly perfect spin polarization of noncollinear antiferromagnets, Nat. Commun. **15**, 10242 (2024).

[18] B. Chi, L. Jiang, Y. Zhu, G. Yu, C. Wan, J. Zhang, and X. Han, Crystal-facet-oriented altermagnets for detecting ferromagnetic and antiferromagnetic states by giant tunneling magnetoresistance, Phys. Rev. Appl. **21**, 34038 (2024).

[19] D.-F. Shao, S.-H. Zhang, M. Li, C.-B. Eom, and E. Y. Tsymbal, Spin-neutral currents for spintronics, Nat. Commun. **12**, 7061 (2021).

[20] Q. Song, S. Stavrić, P. Barone, A. Droghetti, D. S. Antonenko, J. W. F. Venderbos, C. A. Occhialini, B. Ilyas, E. Ergeçen, N. Gedik, et al., Electrical switching of a p-wave magnet, Nature **642**, 64 (2025).

[21] L. M. Corliss, N. Elliott, and J. M. Hastings, Antiferromagnetic structures of $MnS_2$, $MnSe_2$, and $MnTe_2$, J. Appl. Phys. **29**, 391 (1958).

[22] M. Zeng, M.-Y. Zhu, Y.-P. Zhu, X.-R. Liu, X.-M. Ma, Y.-J. Hao, P. Liu, G. Qu, Y. Yang, Z. Jiang, et al., Observation of spin splitting in room-temperature metallic antiferromagnet CrSb, Adv. Sci. **11**, 2406529 (2024).

[23] S. Polesya, G. Kuhn, S. Mankovsky, H. Ebert, M. Regus, and W. Bensch, Structural and magnetic properties of CrSb compounds: NiAs structure, J. Phys.: Condens. Matter **24**, 036004

(2011).

[24] X. Chen, T. Higo, K. Tanaka, T. Nomoto, H. Tsai, H. Idzuchi, M. Shiga, S. Sakamoto, R. Ando, H. Kosaki, et al., Octupole-driven magnetoresistance in an antiferromagnetic tunnel junction, Nature **613**, 7944 (2023).

[25] G. Gurung, D.-F. Shao, and E. Y. Tsymbal, Transport spin polarization of noncollinear antiferromagnetic antiperovskites, Phys. Rev. Mater. **5**, 124411 (2021).

[26] K. Kang, D. G. Cahill, and A. Schleife, Phonon, electron, and magnon excitations in antiferromagnetic $L1_0$-type MnPt, Phys. Rev. B **107**, 064412 (2023).

[27] C. Du, H. Wang, F. Yang, and P. C. Hammel, Systematic variation of spin-orbit coupling with $d$-orbital filling: large inverse spin Hall effect in $3d$ transition metals, Phys. Rev. B **90**, 140407 (2014).

[28] S. Polesya, S. Mankovsky, D. Ködderitzsch, J. Minár, and H. Ebert, Finite-temperature magnetism of FeRh compounds, Phys. Rev. B **93**, 024423 (2016).

[29] I. A. Zhuravlev, A. Adhikari, and K. D. Belashchenko, Perpendicular magnetic anisotropy in bulk and thin-film CuMnAs for antiferromagnetic memory applications, Appl. Phys. Lett. **113**, 162404 (2018).

[30] M. S. Gebre, R. K. Banner, K. Kang, K. Qu, H. Cao, A. Schleife, and D. P. Shoemaker, Magnetic anisotropy in single-crystalline antiferromagnetic $Mn_2Au$, Phys. Rev. Mater. **8**, 84413 (2024).

[31] L. Salemi and P. M. Oppeneer, Theory of magnetic spin and orbital Hall and Nernst effects in bulk ferromagnets, Phys. Rev. B **106**, 024410 (2022).

[32] Y. Zhang, Q. Xu, K. Koepernik, R. Rezaev, O. Janson, J. Železný, T. Jungwirth, C. Felser, J. van den Brink, and Y. Sun, Different types of spin currents in the comprehensive materials database of nonmagnetic spin Hall effect, Npj Comput. Mater. **7**, 1 (2021).

[33] H. Fujii, Y. Uwatoko, K. Motoya, Y. Ito, and T. Okamoto, Neutron diffraction and magnetic studies of CeZn and NdZn single crystals, J. Magn. Magn. Mater. **63–64**, 114 (1987).

[34] E. Krén, G. Kádár, L. Pál, J. Sólyom, P. Szabó, and T. Tarnóczi, Magnetic structures and exchange interactions in the Mn-Pt system, Phys. Rev. **171**, 574 (1968).

[35] D. Solina, W. Schmidt, R. Kaltofen, C. Krien, C.-H. Lai, and A. Schreyer, The magnetic structure of $L1_0$ ordered MnPt at room temperature determined using polarized neutron diffraction, Mater. Res. Express **6**, 76105 (2019).

[36] J. Železný, H. Gao, A. Manchon, F. Freimuth, Y. Mokrousov, J. Zemen, J. Mašek, J. Sinova, and T. Jungwirth, Spin-orbit torques in locally and globally noncentrosymmetric crystals: Antiferromagnets and ferromagnets, Phys. Rev. B **95**, 014403 (2017).

[37] M. Seemann, D. Ködderitzsch, S. Wimmer, and H. Ebert, Symmetry-imposed shape of linear response tensors, Phys. Rev. B **92**, 155138 (2015).

[38] See supplemental material for the computational methods, calculated longitudinal spin conductivity of MnPt and $Mn_2Au$, magnetic point group symmetry analysis of MnPt, and band structures of $Mn_2Au$.

[39] M. Zhu, X. Li, F. Zheng, J. Dong, Y. Zhou, K. Wu, and J. Zhang, Crystal facet orientation and temperature dependence of charge and spin Hall effects in noncollinear antiferromagnets: a

first-principles investigation, Phys. Rev. B **110**, 54420 (2024).

[40] A. Bose, N. J. Schreiber, R. Jain, D.-F. Shao, H. P. Nair, J. Sun, X. S. Zhang, D. A. Muller, E. Y. Tsymbal, D. G. Schlom, et al., Tilted spin current generated by the collinear antiferromagnet ruthenium dioxide, Nat. Electron. **5**, 267 (2022).

[41] P. Wadley, V. Novák, R. P. Campion, C. Rinaldi, X. Martí, H. Reichlová, J. Železný, J. Gazquez, M. A. Roldan, M. Varela, et al., Tetragonal phase of epitaxial room-temperature antiferromagnet CuMnAs, Nat. Commun. **4**, 2322 (2013).

[42] J. Železný, H. Gao, K. Výborný, J. Zemen, J. Mašek, A. Manchon, J. Wunderlich, J. Sinova, and T. Jungwirth, Relativistic Néel-order fields induced by electrical current in antiferromagnets, Phys. Rev. Lett. **113**, 157201 (2014).

[43] P. Wadley, B. Howells, J. Železný, C. Andrews, V. Hills, R. P. Campion, V. Novák, K. Olejník, F. Maccherozzi, S. S. Dhesi, et al., Electrical switching of an antiferromagnet, Science **351**, 587 (2016).

[44] S. Y. Bodnar, L. Šmejkal, I. Turek, T. Jungwirth, O. Gomonay, J. Sinova, A. A. Sapozhnik, H.-J. Elmers, M. Kläui, and M. Jourdan, Writing and reading antiferromagnetic $Mn_2Au$ by Néel spin-orbit torques and large anisotropic magnetoresistance, Nat. Commun. **9**, 348 (2018).

[45] M. H. Fischer, A. Vaezi, A. Manchon, and E.-A. Kim, Spin-torque generation in topological insulator based heterostructures, Phys. Rev. B **93**, 125303 (2016).

[46] I. Mihai Miron, G. Gaudin, S. Auffret, B. Rodmacq, A. Schuhl, S. Pizzini, J. Vogel, and P. Gambardella, Current-driven spin torque induced by the Rashba effect in a ferromagnetic metal layer, Nat. Mater. **9**, 230 (2010).

[47] K. Uchida, S. Takahashi, K. Harii, J. Ieda, W. Koshibae, K. Ando, S. Maekawa, and E. Saitoh, Observation of the spin Seebeck effect, Nature **455**, 778 (2008).